\documentclass[aip,reprint]{revtex4-1}

\usepackage{graphicx}
\usepackage{dcolumn}
\usepackage{bm}
\usepackage{textcomp}
\usepackage{gensymb}
\usepackage[articletitle=true]{achemso}

\draft 

\begin{document}

\preprint{AIP/123-QED}

\title{Ion exchange gels enhance organic electrochemical transistor performance in aqueous solution}

\author{Connor G. Bischak}
\affiliation{Department of Chemistry, University of Washington, Seattle, Washington 98195-1700, United States}

\author{Lucas Q. Flagg}
\affiliation{Department of Chemistry, University of Washington, Seattle, Washington 98195-1700, United States}

\author{David S. Ginger}
\altaffiliation{Corresponding Author (dginger@uw.edu)}
\affiliation{Department of Chemistry, University of Washington, Seattle, Washington 98195-1700, United States}

\begin{abstract}
Conjugated polymer-based organic electrochemical transistors (OECTs) are being studied for applications ranging from biochemical sensing to neural interfaces. While new conjugated polymers are being developed that can interface digital electronics with the aqueous chemistry of life, the vast majority of high-performance, high-mobility organic transistor materials developed over the past decades are extremely poor at taking up biologically-relevant ions. Here we incorporate an ion exchange gel into an OECT demonstrating that this structure is capable of taking up biologically-relevant ions from solution and injecting larger, more hydrophobic ions into the underlying polymer semiconductor active layer in multiple hydrophobic conjugated polymers. Using poly[2,5-bis(3-tetradecylthiophen-2-yl) thieno[3,2-b]thiophene] (PBTTT) as a model semiconductor active layer and a blend of the ionic liquid 1-butyl-3-methylimidazolium bis(trifluoromethylsulfonyl)imide (BMIM TFSI) and poly(vinylidene fluoride-co-hexafluoropropylene) (PVDF-HFP) as the ion exchange gel, we demonstrate more than a four order of magnitude improvement in OECT device transconductance and a one hundred-fold increase in ion injection kinetics. We demonstrate the ability of the ion exchange gel OECT to record biological signals by measuring the action potentials of a Venus flytrap plant. These results show the possibility of using interface engineering to open up a wider palette of organic semiconductor materials as OECTs that can be gated by aqueous solutions.

\end{abstract}


\maketitle 
Because of their ability to conduct both electronic and ionic charge carriers, conjugated polymers have emerged as promising materials for use in a wide array of applications, such as energy storage devices,\cite{Milczarek2012Renewable,lopez_designing_2019,Moia2019Design} neuromorphic computing,\cite{Burgt2017non-volatile,fuller_parallel_2019}
and bioelectronics.\cite{Berggren2007Organic,Someya2016rise,Inal2018Conjugated,Sjostrom2018Decade} One technology that exploits the dual electronic/ionic conductivity of conjugated polymers for bioelectronics is the organic electrochemical transistor (OECT),\cite{Rivnay2018Organic} which has been used for many applications,\cite{Strakosas2015organic} including recording electrical impulses in humans and cell populations,\cite{Campana2014Electrocardiographic,Yao2015Rigid,Gu201616-Channel} \textit{in vivo} electrophysiological recordings,\cite{Khodagholy2013In} \textit{in vitro} biosensing,\cite{Shim2009All-Plastic,Yang2010Electrochemical,Sessolo2013Easy-to-Fabricate,Macchia2018Ultra-sensitive} and neuronal stimulation.\cite{Williamson2015Localized} A distinguishing feature of OECTs compared to conventional organic field-effect transistors (OFETs), is that upon application of an electrochemical gate bias, ions from the electrolyte directly enter the polymer to charge compensate the electronic carriers injected into the organic semiconductors, thus permitting injection of much higher volumetric carrier densities than would be possible with only dielectric surface gating.

To achieve OECTs with high transconductance, the conjugated polymer active layer should have both a high electron or hole mobility ($\mu$) and a high volumetric capacitance ($C$*).\cite{Inal2017Benchmarking} A high $\mu$ is achieved through a tightly-packed, often crystalline, polymer morphology, whereas a high $C$* relies on facile ion migration within the polymer.\cite{Giridharagopal2017Electrochemical,Giovannitti2016Controlling} These parameters depend on a number of properties of the conjugated polymer active layer and electrolyte, including polymer chemistry,\cite{Giovannitti2016Controlling,Nielsen2016Molecular,flagg_polymer_2019} polymer microstructure,\cite{Giridharagopal2017Electrochemical,SavvaSolvent} choice of electrolyte anion and cation,\cite{Flagg2018Anion-Dependent,Cendra2019Role} and electrolyte concentration.\cite{Savva2019Influence} For OECT applications, ion injection kinetics are also an important figure-of-merit: sub-second device switching speeds are desirable for capturing transient biological signals, while even faster kinetics might be beneficial for electrophysiology. OECT kinetics can depend on the degree of conformational changes of the polymer upon electrochemical doping, as well as the chemistry of the polymer sidechains.\cite{Giovannitti2016Controlling,flagg_polymer_2019} Unfortunately, most p-type semiconducting polymers traditionally used for OFETs and non-aqueous OECTs are hydrophobic and unable to take up and transport biologically-relevant ions, functioning poorly as OECT active layers in aqueous electrolytes.\cite{Giovannitti2016Controlling,flagg_polymer_2019} 

\begin{figure*}
\includegraphics[width=14cm]{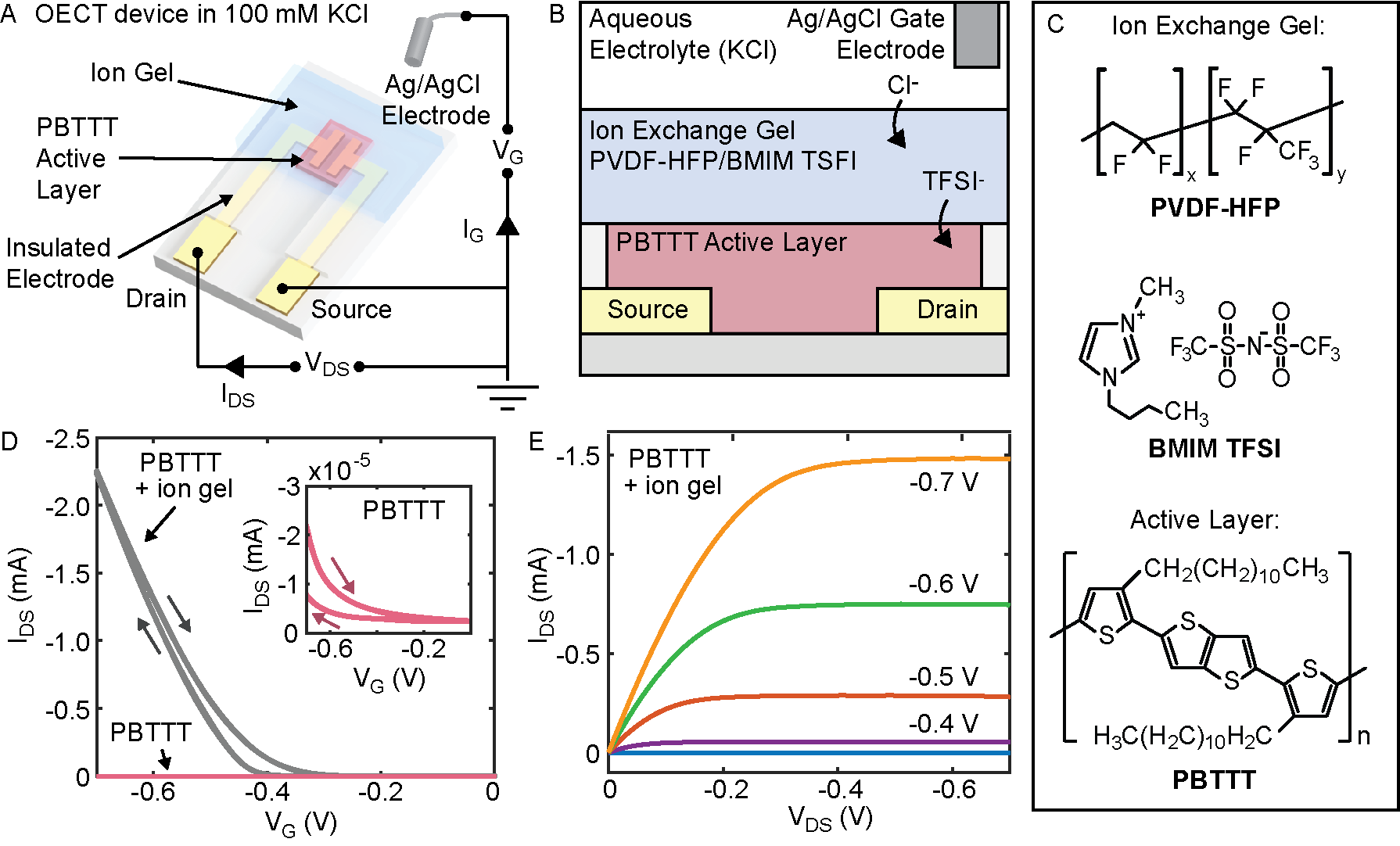}
\caption{\label{fig1}Schematic of OECT with ion exchange gel and transistor evaluation with and without the ion exchange gel. (A) OECT device schematic with the ion exchange gel positioned on top of the active layer. Upon application of a gate bias, ions from solution (Cl\textsuperscript{-}) enter the ion exchange gel and ions from the ion exchange gel (TFSI\textsuperscript{-}) enter and dope the semiconductor polymer active layer. (B) Schematic of OECT operation with the ion exchange gel and p-type active layer. (C) Chemical structures of the polymer semiconductor active layer, PBTTT, and the ion exchange gel, comprising PVDF-HFP and BMIM TFSI ionic liquid. (D) Transfer curve of a PBTTT OECT with and without the ion exchange gel (\textit{V\textsubscript{DS}} = -0.6 V, \textit{d} = 55 nm, \textit{L} = 20 $\mu$m, W = 100 $\mu$m). The inset shows the transfer curve of a PBTTT OECT without the ion exchange gel, which has five orders of magnitude lower operating current, and significantly more hysteresis. The $\mu$$C$* of PBTTT with the ion gel is 179 $\pm$ 40 F cm\textsuperscript{-1} V\textsuperscript{-1} s\textsuperscript{-1} as determined by a series of transistor measurements with different \textit{L} and \textit{d}. (E) Output curves (V\textsubscript{G} = 0.3-0.7 V) for the PBTTT OECT with the ion exchange gel.}
\end{figure*}

A number of promising strategies have been pursued to improve the performance of conjugated polymers for OECT active layers. OECTs with non-aqueous gate dielectrics, such as ionic liquid-based ion gels\cite{Lee2007Ion,Cho2008Printable,Lee2012“Cut} and polymeric ionic liquids,\cite{Hamedi2009Fiber-Embedded,Choi2015Single} operate using hydrophobic conjugated polymer active layers, yet these polymers are generally incompatible with aqueous gate dielectrics, precluding their use in bioelectronics or biosensing applications. To enhance ion transport in conjugated polymers, many strategies, including developing new device designs\cite{Duong2018Universal} and new conjugated polymers, have been pursued. Recently, conjugated polymers and small molecules with ethylene glycol-based or charged side chains have been used for OECTs, including both p-type\cite{Giovannitti2016Controlling,flagg_polymer_2019,GiovannittiRedox-Stability,Savagian2018Balancing,ParrGlycolated,Schmode2019High-Performance} and n-type organic semiconductors.\cite{Giovannitti2016N-type,Giovannitti2018Role,Sun2018n-Type,Bischak2019Fullerene} Unlike polymers with alkyl side chains, these more hydrophilic polymers are better at transporting biologically-relevant ions, such as Cl\textsuperscript{-}, K\textsuperscript{+}, and Na\textsuperscript{+}. On the other hand, hydrophobic polymers are capable of transporting larger, more hydrophobic ions, such as hexafluorophosphate (PF\textsubscript{6}\textsuperscript{-}) or bis(trifluoromethanesulfonyl)imide (TFSI\textsuperscript{-}),\cite{Flagg2018Anion-Dependent,Cendra2019Role} probably because these larger ions are more polarizable, and bind weakly to surrounding water molecules, lowering the threshold voltage for ion injection and enabling transport within the hydrophobic polymer matrix. These larger, hydrophobic ions, however, are not commonly found in nature, so it has been difficult to exploit their favorable properties for bioelectronics.

Herein, we demonstrate a platform that uses an interfacial ionic liquid-based ion exchange gel to enhance the transconductance and ion injection kinetics of OECTs gated with aqueous solutions. We show that the ion exchange gel can significantly improve both the OECT transconductance and kinetics of ion injection for OECTs with active layers composed of a variety of different hydrophobic polymers, including poly[2,5-bis(3-tetradecylthiophen-2-yl) thieno[3,2-b]thiophene] (PBTTT), a common polymer previously used in OFETs\cite{McCulloch2006Liquid-crystalline} and non-aqueous electrolyte-gated OECTs.\cite{Yuen2007Electrochemical} We evaluate the stability of this device architecture, and show that only polymer regions in contact with the ion exchange gel exhibit enhanced electrochemical doping. Finally, we demonstrate the viability of this OECT device geometry by recording the action potential of a live Venus flytrap plant. Our results demonstrate a promising new architecture for OECTs that exchanges biologically-relevant ions for large, hydrophobic ions, resulting in vastly enhanced steady-state doping levels and faster ion injection kinetics in commonly-available, hydrophobic conjugated polymers.

\begin{figure*}
\includegraphics[width=14cm]{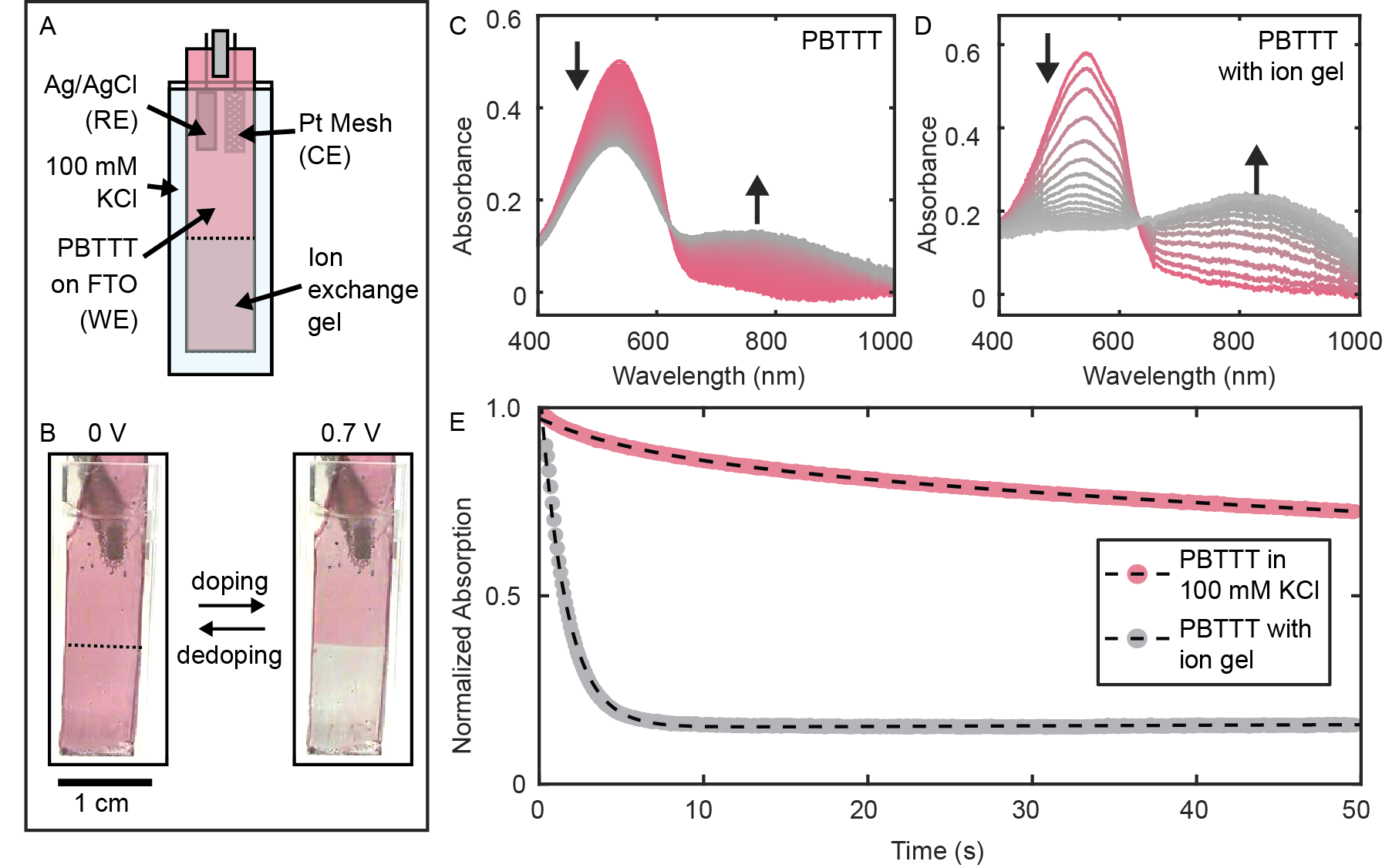}
\caption{\label{fig2} The ion exchange gel improves ion injection kinetics. Spectroelectrochemistry of PBTTT (A) without and (B) with the ion exchange gel, and P3HT (C) without and (D) with the ion exchange gel recorded in 100 mM KCl. Each spectrum is acquired with a 100 ms integration time with 300 ms between spectra (E) Kinetics traces of the decrease in the absorption peaks as a function of time with biexponential fits   (black solid and dotted lines). When applying the ion exchange gel, the change in absorption upon doping increases and the time constant for ion injection decreases. For PBTTT, the predominant decay constant decreases from 68.8 $\pm$ 1.1 s to 1.53 $\pm$ 0.02 s when the ion gel is applied.}
\end{figure*}

Figure 1A shows the device architecture of an OECT incorporating an ion exchange gel between the conjugated polymer active layer and the aqueous electrolyte. Upon application of a gate bias (V\textsubscript{G}), Cl\textsuperscript{-} ions from the aqueous electrolyte enter the ion exchange gel and TFSI\textsuperscript{-} ions enter the underlying semiconductor polymer active layer, compensating for holes on the polymer backbone (Figure 1B). The ion exchange gel is composed of a blend of poly(vinylidene fluoride-co-hexafluoropropylene) (PVDF-HFP) as the gel matrix and 1-butyl-3-methylimidazolium bis(trifluoromethylsulfonyl)imide (BMIM TFSI) as the ionic liquid (Figure 1C). We choose BMIM TFSI as the ionic liquid component because it is immiscible with water and contains TFSI\textsuperscript{-}, an anion that has been previously shown to efficiently dope and diffuse within hydrophobic conjugated polymers.\cite{flagg_polymer_2019,Lee2009Ion} We found that other ionic liquids with smaller, more hydrophilic anions worked, but require a higher bias to dope the underlying hydrophobic polymer (Figure S1). We focus on characterizing this device geometry using PBTTT as the semiconductor polymer active layer (Figure 1C). PBTTT has been used widely as the active layer in OFETs\cite{Allard2008Organic} and has a relatively high charge carrier mobility of $\sim$1.0 cm\textsuperscript{2} V\textsuperscript{-1}  s\textsuperscript{-1},\cite{McCulloch2006Liquid-crystalline} yet this polymer has hydrophobic alkyl side chains and appears unable to uptake biologically-relevant anions in the absence of the ion exchange gel (Figure 1D).

We compare the performance of OECTs with and without the ion gel present using PBTTT as the conjugated polymer active layer. Figure 1D shows transfer curves of an OECT with a PBTTT active layer with and without the ion gel. Figure 1E shows the corresponding output curves with the ion exchange gel applied. The transfer curve shows an increase in the operating current at V\textsubscript{G}  = -0.7 V of more than four orders of magnitude upon adding the ion exchange gel. The transconductance increases from 1.0 x 10\textsuperscript{-4} $\pm$ 9.0 x 10\textsuperscript{-5} mS to 9.0 $\pm$ 1.8 mS upon applying the ion exchange gel on devices with the same polymer film thickness (d = 55 nm) and device dimensions (L = 20 $\mu$m, W = 100 $\mu$m). We find that this improvement in OECT performance when gated with aqueous electrolytes is a general phenomenon for hydrophobic conjugated polymers when using the ion exchange gel: OECTs with a poly-3-hexylthiophene (P3HT) active layer exhibit a $\sim$24,000-fold improvement in the transconductance from 1.2 x 10\textsuperscript{-4} ± 1.1 x 10\textsuperscript{-4} mS to 2.9 $\pm$ 0.6 mS on the same device   with and without the ion exchange gel (\textit{d} = 80 nm, \textit{L} = 20 $\mu$m, \textit{W} = 100 $\mu$m) (Figure S2). 

To compare the ion-exchange OECT to other state-of-the-art OECTs, we measure the $\mu$$C$* product, an intrinsic property of the polymer active layer that is independent of transistor dimensions and film thickness.\cite{Inal2017Benchmarking} By characterizing many transistors with different dimensions and active layer thicknesses, we calculate a $\mu$$C$* of 179 $\pm$ 40 F cm\textsuperscript{-1} V\textsuperscript{-1} s\textsuperscript{-1} for PBTTT and 63 $\pm$ 18 F cm\textsuperscript{-1} V\textsuperscript{-1} s\textsuperscript{-1} for P3HT with the ion exchange gel applied (Figure S3). Notably, the $\mu$$C$* for PBTTT with the ion exchange gel approaches that of the best-performing accumulation mode OECT materials ($\mu$$C$* = $\sim$250 F cm\textsuperscript{-1} V\textsuperscript{-1}) reported to date.\cite{Inal2017Benchmarking} When used with P3HT, the $\mu$$C$* of the ion exchange gel OECT is on the same order as that of typical PEDOT:PSS-based OECTs ($\mu$$C$* = $\sim$60 F cm\textsuperscript{-1} V\textsuperscript{-1}).\cite{Inal2017Benchmarking} 

\begin{figure*}
\includegraphics[width=14cm]{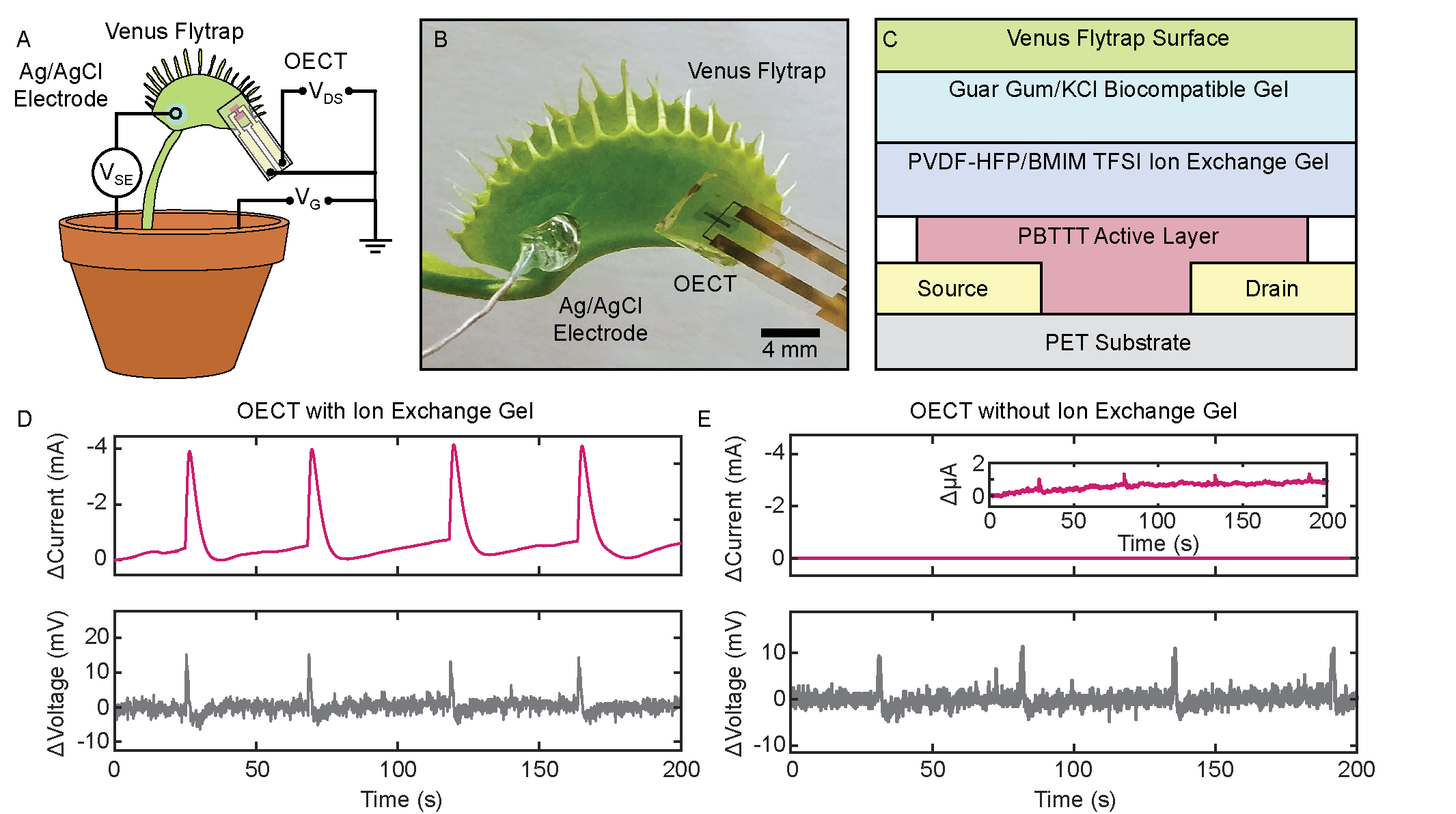}
\caption{\label{fig3}Measuring action potentials in a Venus flytrap using the OECT with ion exchange gel. (A) Schematic of the standard Ag/AgCl electrode and PBTTT-based OECT attached to a Venus flytrap. The gate electrode is placed in the plant's soil. (B) Image of an OECT and Ag/AgCl electrode on the surface on a Venus flytrap trap. (C) Schematic showing the device geometry on the Venus flytrap surface with guar gum/KCl as a biocompatible adhesive gel. (D) Current as a function of time from an OECT and voltage as a function of time from a Ag/AgCl electrode upon triggering the hair inside the flytrap four times. (E) Simultaneous measurement of an OECT and Ag/AgCl electrode with no ion exchange gel. The inset shows change in current versus time with a current scale that captures the signal recorded.}
\end{figure*}

In addition to improving the steady-state OECT device performance, we show that the ion injection rate increases by up to two orders of magnitude when the ion gel is applied. Figure 2A shows a schematic of the spectroelectrochemistry setup used to measure ion injection kinetics. We submerge a PBTTT-coated FTO/glass slide in an electrochemical cell containing 100 mM KCl with a Pt mesh counter electrode (CE) and an Ag/AgCl reference electrode (RE). The FTO-coated glass slide serves as the working electrode (WE). Figure 2B shows images of the ion gel on the polymer thin film before and after application of a bias, demonstrating that enhanced electrochemical doping occurs underneath the ion exchange gel upon applying a bias and minimal doping occurs in areas of the film directly exposed to the KCl electrolyte. Movie S1 and Movie S2 also show reversible enhanced doping under the ion exchange gel in PBTTT and P3HT, respectively, upon performing multiple cyclic voltammograms. To show that this selective doping is a general phenomenon for hydrophobic conjugated polymers, we also demonstrate enhanced doping with the ion exchange gel applied in both PTB7 and PCPDTBT, which can be oxidized in the electrochemical window of water (Figure S4). In the same configuration, we assessed the stability of the ion exchange gel by performing 100 successive cyclic voltammograms. The reversibly injected charge level decreases by only $\sim$10\% after 100 cyclic voltammograms performed on PBTTT film covered with the ion exchange gel, and does not appear to degrade on P3HT films over 100 cycles (Figure S5).  

By using spectroelectrochemistry to measure changes in absorption upon applying a bias, we determine the kinetics of electrochemically-driven ion injection and doping. Figure 2C shows a series of absorption spectra taken following application of a 0.7 V bias (vs. Ag/AgCl) to PBTTT. These spectra show a small decrease in the main absorption peak and an increase of the near-IR polaron absorption peak. Figure 2D shows the results of the same experiment with the ion exchange gel modified PBTTT film, which results in faster doping and a more substantial decrease in the main absorption peak and increase in the polaron absorption. We fit the decrease of the absorption spectrum maximum of PBTTT to a biexponential decay and find that the main component decreases from 68.8 $\pm$ 1.1 s to 1.53 $\pm$ 0.02 s upon adding the ion gel, a $\sim$50x increase in ion injection rate. For P3HT, we find a similar decrease of 152 $\pm$ 12 s to 1.64 $\pm$ 0.02 s, a $\sim$100x increase the ion injection rate when the ion gel is applied (Figure S6). We additionally measured transistor switching speeds by modulating V\textsubscript{G} between 0 and -0.7 V and find a similar time constant for transistor turn on (3.1 $\pm$ 0.1 s) and much faster time constant for transistor turn off (0.14 $\pm$ 0.01 s) (Figure S7). We also evaluated the frequency response of OECTs with and without the ion exchange gel (Figure S8) and, consistent with the device measurements, found that the device does not turn on without the ion exchange gel. We note that while the ion-gel-modified OECTs operate with improved kinetics relative to unmodified OECTs based on hydrophobic polymers, their kinetics are still $\sim$10-100x slower than OECTs containing conjugated polymers with glycol sidechains and PEDOT:PSS-based OECTs.\cite{Giovannitti2016Controlling,flagg_polymer_2019,Rivnay2015High-performance} While second-scale switching speeds may already be sufficient for some biosensing applications, we speculate that it should be possible to improve kinetics and switching speeds further by modifying the ion gel/polymer interface, and optimizing the device geometry and/or polymer morphology.

Finally, as a practical demonstration, we explore the viability of the ion-exchange gel-modified OECTs for measuring extracellular action potentials by using them to record the action potential of a Venus flytrap (\textit{Dionaea muscipula}) upon mechanical stimulation. The Venus flytrap serves as a model system for recording extracellular signals with an action potential that consists of a sharp depolarization peak followed by a longer hyperpolarization.\cite{BurdonSanderson1873.,Darwin1888Insectivorous,Hodick1986influence,Forterre2005How} The upper leaf of the Venus flytrap typically contains six trigger hairs, which generate an action potential when stimulated. The trap closes when two mechanical stimuli occur within 0.75-20 s, placing constraints on the time resolution of the system. We placed both a commercial Ag/AgCl wire electrode and an OECT on the surface of the Venus flytrap leaf and recorded signals from both the OECT and Ag/AgCl electrodes simultaneously while stimulating the trigger hairs to trigger the action potential. We show the positions of the electrodes and circuit diagram in Figure 3A and an image of the electrodes on a Venus flytrap leaf in Figure 3B. Figure 3C shows the device geometry of the OECT with a PBTTT active layer, PVDF-HFP/BMIM TFSI ion exchange gel, and a biocompatible gel composed of a mixture of guar gum and an aqueous 100 mM KCl solution used to interface with the plant and prevent direct contact between the ion exchange gel and the plant. 

We simultaneously record a series of action potentials with the OECT and standard Ag/AgCl electrode by stimulating the trigger hairs on the Venus flytrap repeatedly. Movie S3 and Movie S4 show OECT action potentials recorded upon flytrap stimulation. Traces of voltage versus time show the simultaneous recording of the action potential with both the ion exchange gel PBTTT OECT (\textit{L} = 20 $\mu$m, \textit{W} = 400 $\mu$m, \textit{d} = 55 nm) and a standard commercial Ag/AgCl wire electrode (Figure 3D). The Ag/AgCl electrode serves as a control and measures changes in potential between the Venus flytrap leaf and ground. We calculate the signal-to-noise ratio (SNR) by dividing the amplitude of the depolarization peak (\textit{I\textsubscript{A}} or \textit{V\textsubscript{A}}) by the standard deviation ($\sigma$) of the noise without flytrap stimulation, as has been used previously to calculate the SNRs of OECTs.\cite{Macchia2018Ultra-sensitive} In comparing the SNR without, and with, the ion gel applied (Figure 4D and Figure 4E), we find that the SNR increases from 10 $\pm$ 2 (\textit{I\textsubscript{A}} = 0.27 $\pm$ 0.06 $\mu$A, $\sigma$ = 28 nA) to 1250 $\pm$ 20 (\textit{I\textsubscript{A}} = 35 $\pm$ 1 $\mu$A, $\sigma$ = 28 nA) when the ion exchange gel is applied. Without the ion gel, the SNR is similar to that measured with the Ag/AgCl electrode (SNR = 11 $\pm$ 1, \textit{V\textsubscript{A}} = 14 $\pm$ 1 mV, $\sigma$ = 1.3 mV). Although the SNR improves significantly with the ion gel, the OECT operation could still benefit from an increase in the response speed, as evidenced by the broadening of the depolarization peak in the OECT channel. Potential strategies to increase the transistor kinetics include optimizing the polymer thickness, increasing the surface area of polymer / ion gel interface through nano- and micro-structuring, or developing new gel materials that exhibit higher ion mobilities. Nevertheless, the ion-exchange gels already allow conversion of existing OFET platforms into aqueous OECTs capable of recording signals in aqueous and biological environments. 

We have used an ion exchange gel to enhance the properties of hydrophobic polymers when incorporated into OECTs in contact with aqueous electrolytes. By exchanging Cl\textsuperscript{-} in the aqueous electrolyte with TFSI\textsuperscript{-} in the ion exchange gel, we show enhanced device performance with transconductance values improving by up to four orders of magnitude and a $\mu$$C$* approaching the highest performing accumulation mode OECT materials. We additionally show an improvement in the ion injection kinetics by as much as 100x and stability of the ion gel under repeated cycling. To show the utility of the ion exchange gel, we measure the action potential of a Venus flytrap with and without the ion exchange gel and show over two orders of magnitude improvement in the SNR upon incorporating the ion exchange gel. Overall, using an ion exchange gel for enhanced OECT performance is a promising, straight-forward means to turn an otherwise   poorly-performing hydrophobic polymer into a high-performing, sensitive OECT active layer material. This technology allows for the fine-tuning of polymer-ion interactions independent of the aqueous electrolyte and provides a promising route for OECT device optimization through interfacial layers.

\section*{Acknowledgments}

C. G. B. is a Washington Research Foundation Postdoctoral Fellow. L. Q. F.'s contribution was based in part on work supported by the expired National Science Foundation award, NSF DMR-1607242, and subsequent support from NWIMPACT SEED award funding.

\section*{Author Contributions}

C. G. B., L. Q. F. and D. S. G. conceived of the experiments and wrote the manuscript. C. G. B. fabricated ion exchange gel and OECTs and performed all transistor measurements and spectroelectrochemistry. C.G.B and L.Q.F. measured action potentials from the Venus flytrap with the OECTs.

\section*{References}

\bibliography{bibtexrefs.bib}

\end{document}